\documentclass[aps,prb,twocolumn,showpacs,psfig,superscriptaddress,longbibliography]{revtex4-2}
\usepackage{listings}
\usepackage{tikz}

\usepackage{amssymb}
\usepackage{algpseudocode}
\usepackage{algorithm}
\usepackage{amsmath}
\usepackage{braket}
\usepackage{booktabs}
\usepackage{chngcntr}
\usepackage{color}
\usepackage{comment}
\usepackage{dsfont}
\usepackage{etoolbox}
\usepackage{epstopdf}
\usepackage[T1]{fontenc}
\usepackage{float}
\usepackage{graphicx}
\usepackage[colorlinks=true,linkcolor=blue,citecolor=blue,
urlcolor=blue]{hyperref}
\usepackage[utf8]{inputenc}
\usepackage{indentfirst}
\usepackage{latexsym,bm,euscript}
\usepackage{mathrsfs}
\usepackage{multirow}
\usepackage{natbib}
\usepackage{soul} 
\usepackage{subfigure}
\usepackage{times}
\usepackage{titlesec}
\usepackage{tikz}
\usepackage{txfonts}
\usepackage{xspace}
\usepackage{xfrac}
\usepackage{chemformula}

\usepackage{mathtools}
\usepackage[vlined, ruled, algo2e, linesnumbered]{algorithm2e}
\usetikzlibrary{shapes}

\def\beqn{\begin{eqnarray}}
\def\eeqn{\end{eqnarray}}
\def\beq{\begin{equation}}
\def\eeq{\end{equation}}

\def\genbox#1#2#3#4#5#6{% #1=0/1, #2=color, #3=shape, #4=raise, #5=width, #6=width/2
\leavevmode\raise#4bp\hbox to#5bp{\vrule height#5bp depth0bp width0bp
\pdfliteral{q .5 w \csname #2COLOR\endcsname\space RG
\csname #3PDF\endcsname{#5}{#6} S Q
\ifx1#1 q \csname #2COLOR\endcsname\space rg 
\csname #3PDF\endcsname{#5}{#6} f Q\fi}\hss}}

\graphicspath{{./Fig/}{./Fig/tst_L06/}{./Fig/tst_L24/}}

\begin{document}

\title{Berezinskii-Kosterlitz-Thouless Quantum Supercriticality in XXZ Heisenberg Spin Chain}

\author{Haoshun Chen}
\thanks{These authors contributed equally to this work.}
\affiliation{Beijing National Laboratory for Condensed Matter Physics, Institute of Physics, Chinese Academy of Sciences, Beijing 100190, China}

\author{Enze Lv}
\thanks{These authors contributed equally to this work.}
\affiliation{Institute of Theoretical Physics, Chinese Academy of Sciences, Beijing 100190, China}
\affiliation{School of Physical Sciences, University of Chinese Academy of Sciences, Beijing 100049, China}

\author{Ning Xi}
\affiliation{Institute of Theoretical Physics, Chinese Academy of Sciences, Beijing 100190, China}

\author{Fei Ye}
% \email{yef@sustech.edu.cn}
\affiliation{Department of Physics, Southern University of Science and Technology, Shenzhen 518055, China}

\author{Wei Li}
\email{w.li@itp.ac.cn}
\affiliation{Institute of Theoretical Physics, Chinese Academy of Sciences, Beijing 100190, China}
\affiliation{School of Physical Sciences, University of Chinese Academy of Sciences, Beijing 100049, China}

%===============   Abstract   =============== 
\begin{abstract} 
Quantum fluctuations can give rise to a singular quantum critical point (QCP) in the ground state, whose influence extends to finite temperatures, forming a quantum critical regime (QCR). Recently, it has been shown that in the quantum Ising model, the symmetry-breaking, longitudinal field can induce a quantum supercritical regime (QSR) emanating from the QCP, which hosts a universally enhanced quantum supercritical magnetocaloric effect (MCE)~\cite{Lv2025,wang2025}. In this paper, we show that the QSR also emerges in the spin-1/2 XXZ model, in both the form of Ising and Berezinskii-Kosterlitz-Thouless (BKT) supercriticality. Using ground-state and finite-temperature tensor-network methods, we investigate quantum supercritical phenomena near a BKT QCP. We reveal a quantum supercritical crossover scaling $T \propto h^{2/3}$ and a Grüneisen ratio scaling $\Gamma_h \propto T^{-3/2}$ for the BKT QCP, which differ from the corresponding Ising supercritical scalings. Nevertheless, we find that the scaling function $\phi_{\Gamma}(x)$ of the singular Grüneisen ratio for both BKT and Ising cases can be approximately described by the same expression $\phi_{\Gamma}(x) \approx x/(1+x^2)$. Our work extends the study of quantum supercritical phenomena from the Ising to the XXZ Heisenberg model, thereby revealing the presence of BKT quantum supercriticality and broadening the scope of quantum supercritical physics.
\end{abstract}

\date{\today}\maketitle

%===============   Introduction   ===============
\section{Introduction}
\label{secI}
In strongly correlated systems, the non-thermal parameter $g$ can induce quantum fluctuations and drive a continuous quantum phase transition --- quantum critical point (QCP) --- in the ground state, giving rise to fertile non-classical phenomena and intriguing behaviors~\cite{Sachdev2000,Coleman2005,sachdev2015}. At the QCP with $g=g_c$, the correlation length diverges and the system becomes scale-invariant, yielding scaling laws and universal behaviors~\cite{sondhi1997,continentino2017,Xiang2025}. Because of these exotic characteristics, quantum criticality becomes the focus of modern condensed matter physics, and has been observed experimentally in abundant systems, including quantum magnets~\cite{He2005,Sebastian2006,Sachdev2008,coldea2010,liang2015,Brando2016,faure2018,Wang2018Nature,wang2019,Li2024TopoCooling,Wang2024Nature,Xiang2024Nature,Xiang2025}, heavy-fermion materials~\cite{Gegenwart2007,Gegenwart2008,Tokiwa2009,Rowley2014,Tokiwa2015Signature,Tokiwa2016Superheavy,Tokiwa2016prl,Gegenwart2016}, and unconventional superconductors~\cite{vanderMarel2003QuantumCB,park2006,hossain2025}. 

\begin{figure*}[t!]
    \centering
    \includegraphics[width=0.65\linewidth]{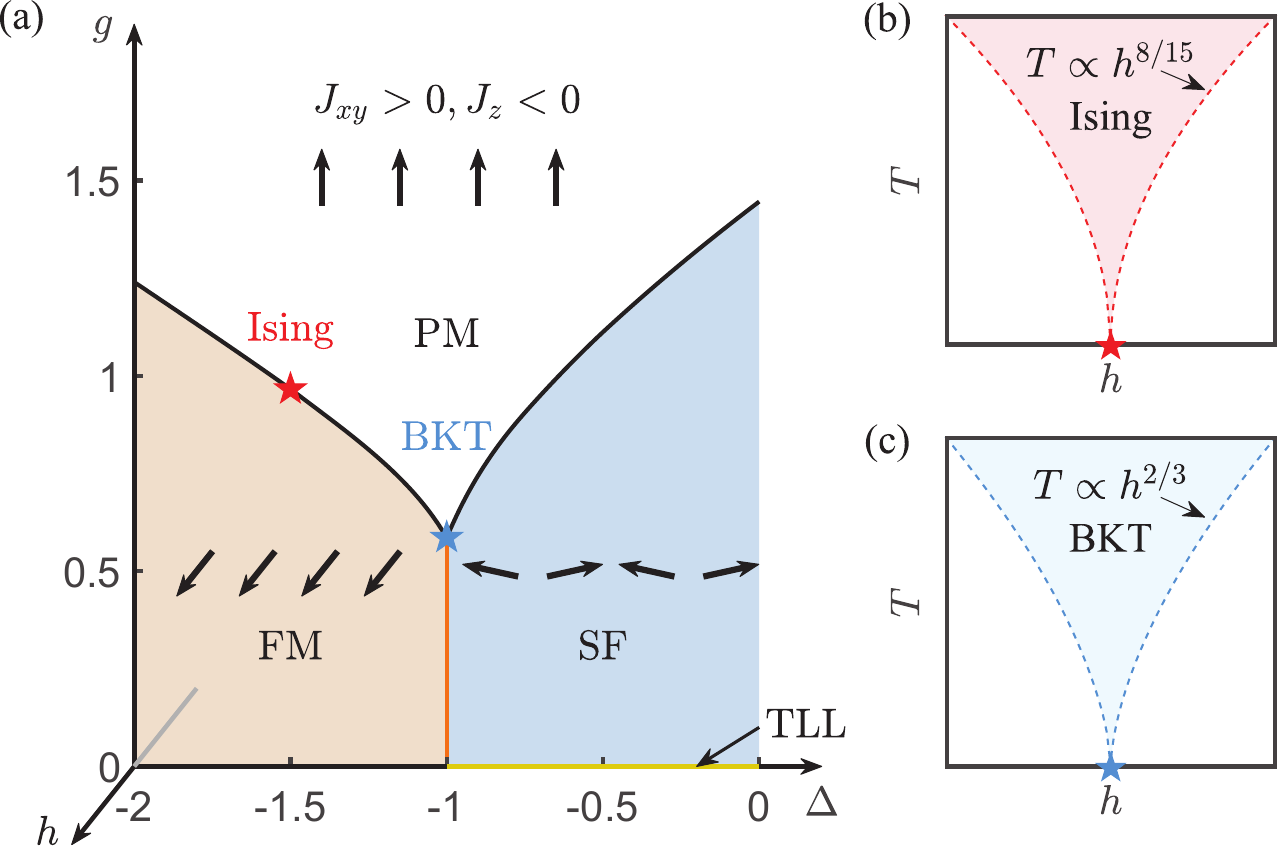}
    \caption{(a) Phase diagram of XXZ spin chain under the longitudinal ($h$) and transverse ($g$) fields. Vertical orange line represents the first-order transition line between the ferromagnetic (FM) and spin-flop (SF) phase. Black curve represents the second-order transition to the paramagnetic (PM) phase. The semitransparent orange plane inside FM phase identifies a first-order transition plane induced by $h$. Red star represents the Ising quantum critical point (QCP), while blue star represents the Berezinskii-Kosterlitz-Thouless (BKT) QCP. (b,c) Quantum supercritical regimes of the Ising QCP (red star) and the BKT QCP (blue star) in $h$-$T$ plane, respectively. Dashed lines represent the quantum supercritical crossover lines with universal scaling $T\propto h^{1/\sigma}$ ($\sigma = 15/8$ for the Ising QCP and $\sigma = 3/2$ for the BKT QCP). }
    \label{fig:Fig1}
\end{figure*}

At finite temperatures, originating from QCP, a cone-shaped quantum critical regime (QCR) emerges in the $g$-$T$ plane, which separates an ordered phase and a quantum disordered phase~\cite{Sachdev2000,Coleman2005,Gegenwart2008}. In this regime, the system exhibits strong quantum critical fluctuations and temperature-dependent scaling laws~\cite{Sachdev2000,Coleman2005,Gegenwart2008,Xiang2025}. Particularly, the crossover lines that enclose the QCR follow a quantum-critical scaling $T\propto \tilde{g}^{z\nu}$, where $\tilde{g}\equiv (g-g_c)/g_c$ measures the distance to the QCP, $z$ is the dynamic critical exponent and $\nu$ is the critical exponent of correlation length. 

Besides the quantum-fluctuation field $g$, a symmetry-breaking field $h$, which couples to the order parameter and breaks the symmetry of Hamiltonian, can give rise to a quantum supercritical regime (QSR) originating from the same QCP in the $h$-$T$ plane~\cite{wang2025,Lv2025}. Unlike QCR, QSR separates two ordered phases and possesses new quantum supercritical scaling laws. Remarkably, the crossover lines of QSR satisfy a new scaling law $T\propto h^{1/\sigma}$, where $\sigma$ is the quantum supercritical exponent. For most universality classes, this exponent reads $\sigma = (\beta+\gamma)/z\nu$~\cite{Lv2025}, with $\beta$ and $\gamma$ the critical exponents of the order parameter and susceptibility, respectively. Compared to QCR, QSR induced by a symmetry-breaking field exhibits a range of superior field-dependent effects, among which the magnetocaloric effect (MCE) stands out as particularly significant~\cite{Weiss1917}. 

At low temperatures, gapless quantum critical excitations lead to a power-law entropy, which is significantly larger than that in a gapped phase. This leads to a pronounced temperature response during an adiabatic demagnetization process. The magnetic Gr\"uneisen ratio --- a key quantity characterizing the MCE --- exhibits a temperature-dependent divergence $\Gamma_g \equiv 1/T(\partial T/ \partial g)_S \propto T^{-1/z\nu}$ near a QCP~\cite{zhu2003, Zhitomirsky2004, Markus2005, Honecker2009, Wu2011JPhCS, zhang2019}. This quantum critical enhanced MCE and associated $T^{-1/z\nu}$ scaling have been observed in many correlated materials~\cite{Wolf2011, Wolf2014, Tokiwa2009, Tokiwa2015Signature, Gegenwart2016, Wolf2016, Oliver2017, Xiang2025}. Remarkably, compared to the quantum-fluctuation field $g$, the symmetry-breaking field $h$ can drive a universally enhanced MCE with the Gr\"uneisen ratio obeying a quantum supercritical scaling $\Gamma_h \equiv 1/T (\partial T/\partial h)_S \propto T^{-(\beta+\gamma)/z\nu}$~\cite{Lv2025}. Typical universality classes host $\beta+\gamma>1$, indicating a universal quantum supercritical boost of the Gr\"uneisen ratio $\Gamma_h \propto \Gamma_g^{\beta+\gamma}$. Therefore, quantum supercriticality with its significant MCE can promote the understanding of quantum scaling laws and inspire the next-generation solid-state refrigeration at ultralow temperatures. 

Here, we study an XXZ spin-1/2 chain with antiferromagnetic (AFM) in-plane and ferromagnetic (FM) out-of-plane interactions. Under both the transverse field $g$ and the longitudinal field $h$, we reveal a Berezinskii-Kosterlitz-Thouless (BKT) QCP, whose quantum supercritical scaling laws are distinct with the Ising case. Then, we investigate the MCE near the BKT QCP and obtain the scaling function $\phi_{\Gamma}(x)$ of the Gr\"uneisen ratio by data collapsing. Remarkably, we propose a concise approximation of the scaling function $\phi_{\Gamma}(x) \approx x/(1+x^2)$, which also applies to the Ising case. In Sec.~\ref{secII}, we introduce the FM-AFM XXZ spin-1/2 chain model and present the ground-state transverse field-anisotropy ($g$-$\Delta$) phase diagram, which hosts a BKT QCP. In Sec.~\ref{secIII}, we exhibit finite-temperature scaling laws of response functions, including longitudinal magnetic susceptibility, transverse magnetic susceptibility and specific heat. In Sec.~\ref{secIV}, we study the quantum supercritical MCE with a diverging Gr\"uneisen ratio, and propose a concise approximation of its quantum supercritical scaling function.

\section{Ground-state phase diagram of FM-AFM XXZ chain}
\label{secII}

\begin{figure*}[t]
    \centering
    \includegraphics[width=0.8\linewidth]{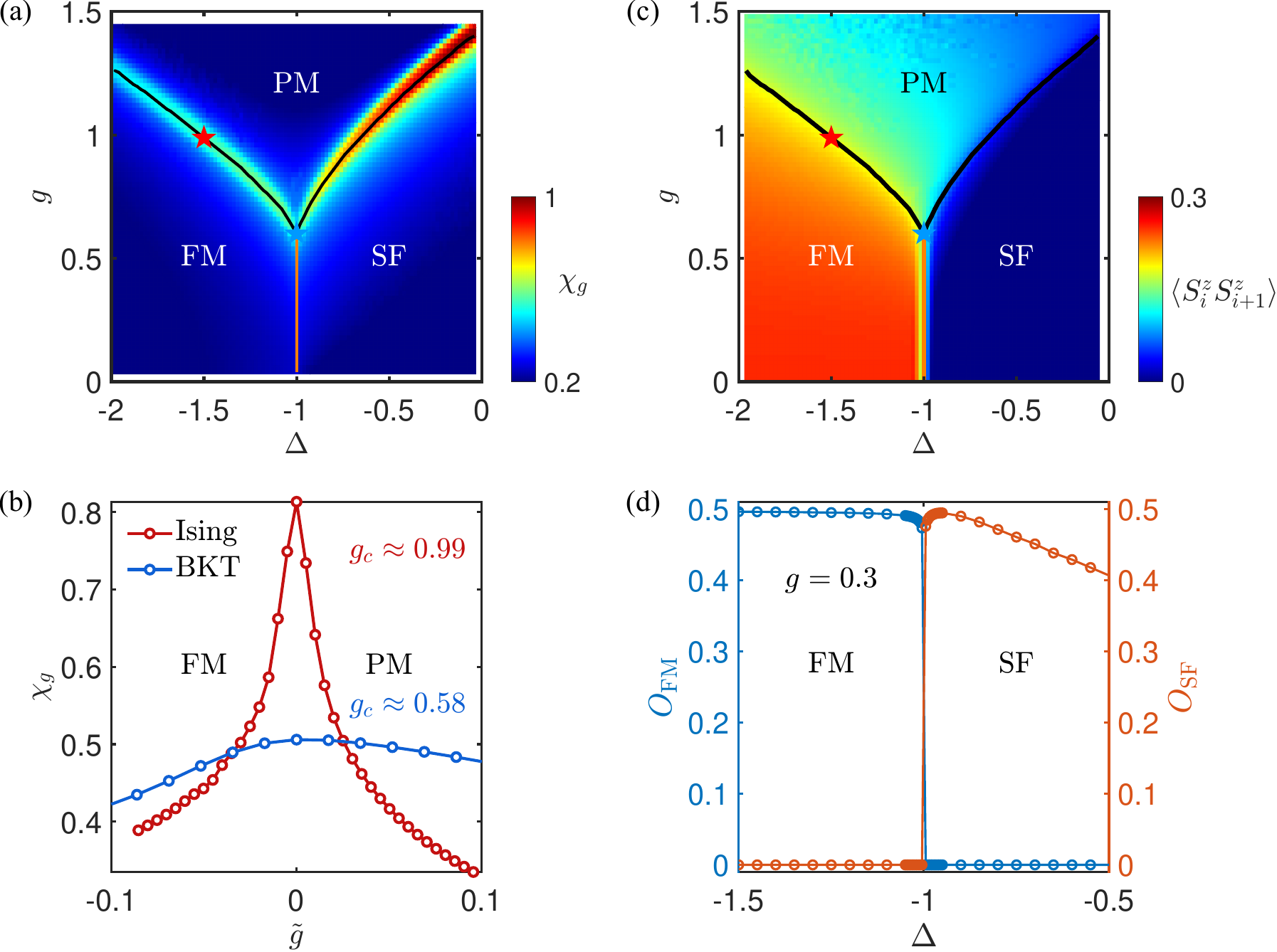}
    \caption{(a) Contour plot of the transverse susceptibility $\chi_g \equiv \partial M_x/\partial g$. The black solid line indicates the location of the maxima in $\chi_g$, representing the second-order Ising transition line. (b) $\chi_g$ as a function of $\tilde{g} \equiv (g-g_c)/g_c$. The red curve corresponds to the case with $\Delta = -1.5$, which passes through the Ising QCP at $g_c \approx 0.99$. The blue curve represents the case with $\Delta = -1$, crossing the BKT QCP at $g_c \approx 0.58$. (c) Contour plot of the nearest-neighbor correlation $\braket{S_{i}^z S_{i+1}^z}$. The orange line is the first-order transition line. (d) Order parameters $O_{\rm FM} \equiv \frac{1}{L}\sum_{i=1}^L\braket{S_i^z}$ and $O_{\rm SF} \equiv \frac{1}{L}\sum_{i=1}^L (-1)^i\braket{S_i^y}$ as a function of $\Delta$ at $g=0.3$.}
    \label{fig:Fig2}
\end{figure*}

In this work, we study the XXZ spin-1/2 chain, whose Hamiltonian reads, 
\begin{equation}
\begin{split}
H & = \sum_i [J_{xy} (S_i^x S_{i+1}^x + S_i^y S_{i+1}^y) + J_z S_i^z S_{i+1}^z ] \\
  & - g\sum_i S_i^x - h\sum_i S_i^z. 
\end{split}
\label{Ham1}
\end{equation}
Here, $J_{xy} \equiv 1$ sets the AFM in-plane coupling as the energy scale, $J_z<0$ denotes the out-of-plane FM interaction, and $\Delta \equiv J_z/J_{xy}$ represents the anisotropy. $g$ is the transverse field that induces quantum fluctuations, while $h$ is the longitudinal field, which couples to the FM order parameter. 

Figure~\ref{fig:Fig1}(a) illustrates the phase diagram of the FM-AFM spin chain system, which hosts fruitful quantum phase transitions driven by the $g$ field in the $h=0$ plane. For the $-1<\Delta<0$ case, the system is in the Tomonaga-Luttinger liquid (TLL) phase with gapless spinon excitations. The transverse field $g$ opens an energy gap and the system enters a gapped spin-flop (SF) phase, where the spins form a long-range AFM order along the $y$ direction. Eventually, the $g$ field drives a continuous quantum phase transition from the SF phase to the paramagnetic (PM) phase, where the spins align along the $x$ axis. This transition line belongs to the (1+1)D Ising universality class due to the $\mathbb{Z}_2$ symmetry of the Hamiltonian. For the $\Delta<-1$ case, the ground state is the FM state, where the spins align along the $z$ direction. In this region, the $g$ field drives a quantum phase transition from the FM to the PM phase. Because of the same symmetry of the Hamiltonian, the FM-PM and the SF-PM quantum phase transitions belong to the same universality class. 

However, $\Delta=-1$ is different. In this case, we can perform a unitary transformation on the Hamiltonian by rotating the spins on even sites by $\pi$ about the $z$ axis, which sends 
\begin{equation}
    S_{2i}^x \rightarrow -S_{2i}^x ,\quad S_{2i}^y \rightarrow -S_{2i}^y.
\end{equation} 
Then, the original FM-AFM chain is equivalent to an FM Heisenberg chain in a staggered transverse field 
\begin{equation}
    H = -J_{xy} \sum_i S_i\cdot S_{i+1} - g\sum_i (-1)^i S_i^x - h\sum_i S_i^z.
\label{Ham2}
\end{equation}
At zero longitudinal field ($h=0$), the system possesses a U(1) symmetry. Consequently, a distinct QCP exists at the critical field $g_c$, which belongs to the BKT universality class. For $g < g_c$, the system is gapless with linear dispersion and exhibits algebraic correlation behaviors (see Appendix for details). Meanwhile, tuning the anisotropy across $\Delta = -1$ induces a first-order transition from the FM phase to the SF phase. 

Using the density matrix renormalization group (DMRG) method~\cite{white1992,itensor}, we calculate the ground-state phase diagram of the FM-AFM XXZ model (see Eq.~\ref{Ham1}). We perform the DMRG calculation on a $L=128$ spin chain under periodic boundary condition in Fig.~\ref{fig:Fig2}(a,c,d) with the maximum bond dimension $D=500$. Particularly, we extend the system size to $L=256$ in Fig.~\ref{fig:Fig2}(b). The truncation error near the QCP is less than $10^{-8}$. 

Figure~\ref{fig:Fig2}(a) illustrates the transverse susceptibility $\chi_g \equiv \partial M_x/\partial g$ in the $g$-$\Delta$ plane. There exist two ``ridges'' of $\chi_g$, revealing strong transverse fluctuations. Specifically, for $\Delta = -1.5$, $\chi_g$ exhibits a sharp peak at the critical field $g_c \approx 0.99$, precisely locating the Ising quantum phase transition [see Fig.~\ref{fig:Fig2}(b)]. Moreover, the quantum phase transition line [black line in Fig.~\ref{fig:Fig2}(a)] is determined by tracking the sharp peaks of $\chi_g$. In contrast, for $\Delta = -1$, $\chi_g$ shows a broad hump rather than a sharp peak, indicating the infinite-order BKT transition at the critical field $g_c\approx 0.58$ (see Appendix for details). 

\begin{figure*}[t!]
    \centering
    \includegraphics[width=1\linewidth]{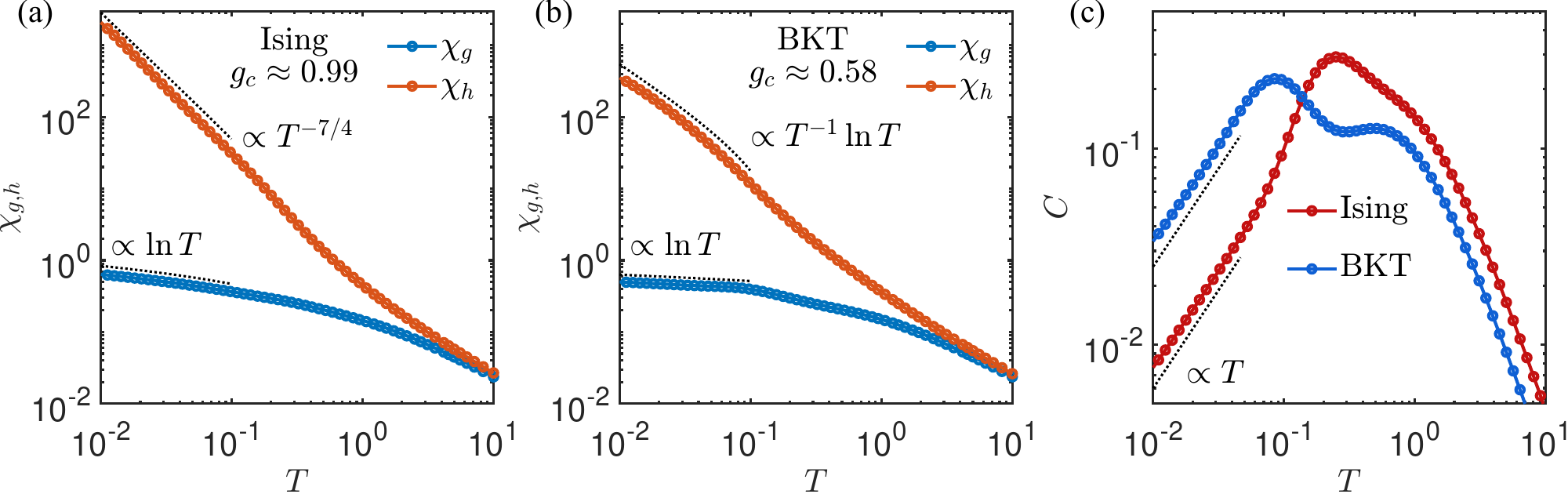}
    \caption{(a,b) Temperature dependence of the longitudinal and transverse magnetic susceptibilities, $\chi_{h}$ and $\chi_{g}$, at the Ising and BKT QCP, respectively. Dashed lines illustrate the low-temperature scaling laws. (c) Specific heat as a function of temperature at these QCPs. Dashed lines indicate the linear behavior at low temperatures.}
    \label{fig:Fig3}
\end{figure*}

However, $\chi_g$ cannot identify the first-order quantum phase transition, as it runs parallel to the $g$ axis. In Fig.~\ref{fig:Fig2}(c), we present a contour plot of the nearest-neighbor correlation $\braket{S_{i}^z S_{i+1}^z}$, which is equivalent to the first derivative $\partial E/\partial \Delta$. This quantity clearly distinguishes these two phases: $\partial E/\partial \Delta > 0$ in the FM phase, whereas $\partial E/\partial \Delta < 0$ in the SF phase. At the first-order transition line, $\partial E/\partial \Delta$ exhibits a discontinuity, jumping from about -0.1 to 0.3. Moreover, we also characterize the first-order phase transition using order parameters: $O_{\rm FM} \equiv \frac{1}{L}\sum_{i=1}^L\braket{S_i^z}$ for the FM phase and $O_{\rm SF} \equiv \frac{1}{L}\sum_{i=1}^L (-1)^i\braket{S_i^y}$ for the SF phase. As illustrated in Fig.~\ref{fig:Fig2}(d), both $O_{\rm FM}$ and $O_{\rm SF}$ exhibit a sudden jump at $\Delta = -1$, providing strong evidence of a first-order phase transition. 

\section{Finite-temperature scaling of response functions}
\label{secIII}

In this section, to calculate the finite-temperature properties~\cite{li2011,Dong2017,Chen2018XTRG,tanTRG2023}, we conduct thermal tensor-network method --- linear tensor renormalization group (LTRG) method on a infinite-size spin system~\cite{li2011,Dong2017}. In our LTRG calculations, the maximum bond dimension is $D = 200$ and the step length of imaginary time evolution is $\tau = 0.01$. 

\subsection{Ising QCP}

Below, we investigate the effect of the longitudinal field $h$ and associated finite-temperature scaling laws. In the vicinity of a QCP, due to scale invariance, the singular part of the free energy $F$ obeys a universal form 
\begin{equation}
    F = T^{d/z+1}\phi_f\left(\frac{h}{T^{\sigma}}, \frac{\tilde{g}}{T^{1/z\nu}}\right), 
    \label{freeenergy}
\end{equation}
where $d$ is the spatial dimension, $\phi_f(x,y)$ with $x\equiv hT^{-\sigma}$ and $y\equiv \tilde{g}T^{-1/z\nu}$ is the hyperscaling function of the free energy. It is worth mentioning that the temperature dependence in $x$ and $y$ differs, leading to distinct finite-temperature behaviors of the longitudinal susceptibility $\chi_{h}$ and the transverse susceptibility $\chi_g$. According to the universal form of the free energy in Eq.~\ref{freeenergy}, the singular parts of these two susceptibilities satisfy
\begin{equation}
    \begin{split}
        & \chi_g \equiv - \frac{\partial^2 F}{\partial g^2} = T^{d/z+1-2/z\nu} g_c^{-2} \partial_y^2\phi_{f}\left(\frac{h}{T^{\sigma}}, \frac{\tilde{g}}{T^{1/z\nu}}\right), \\
        & \chi_h \equiv - \frac{\partial^2 F}{\partial h^2} = T^{d/z+1-2\sigma}\partial_x^2\phi_{f}\left(\frac{h}{T^{\sigma}}, \frac{\tilde{g}}{T^{1/z\nu}}\right), 
    \end{split}
\end{equation}
where $\partial_x^2\phi_f(x,y)$ and $\partial_y^2\phi_f(x,y)$ are the second partial derivative of the scaling function $\phi_f(x,y)$. Using the Josephson scaling law $(d+z)\nu - 2 = -\alpha$ and the Rushbrooke scaling law $\alpha + 2\beta + \gamma = 2$, along with $\sigma = (\beta+\gamma)/z\nu$ for conventional universality classes, we can obtain the temperature-dependent scaling of these susceptibilities, 
\begin{equation}
    \chi_g \propto T^{-\alpha/z\nu}, \quad \chi_h \propto T^{-\gamma/z\nu}.
\end{equation}
Noted that, compared with a finite-temperature critical point with the susceptibility scaling $\chi_h \propto (T-T_c)^{-\gamma}$, the QCP hosts a distinct quantum supercritical scaling $\chi_h \propto T^{-\gamma/z\nu}$, featuring an extra quantum correction $z\nu$.  

In the phase diagram shown in Fig.~\ref{fig:Fig2}, the system exhibits an Ising QCP at $\Delta = -1.5$ and $g_c\approx 0.99$, and the low-energy properties can be effectively described by a $\phi^4$ field theory near this point~\cite{fradkin2013field}. The longitudinal field $h$ couples to the FM order parameter and breaks the $\mathbb{Z}_2$ symmetry, thus driving quantum supercritical phenomena. At finite temperatures, a QSR emerges, as shown in Fig.~\ref{fig:Fig1}(b). The two crossover lines enclosing this regime follow a universal scaling $T\propto h^{1/\sigma}$, where $\sigma = (\beta+\gamma)/z\nu$ is the quantum supercritical exponent~\cite{Lv2025}. For the (1+1)D Ising universality class, the critical exponents are $\beta = 1/8$, $\gamma = 7/4$, $z=1$, and $\nu=1$; thus $\sigma = 15/8$, giving the crossover scaling $T \propto h^{8/15}$~\cite{Di1997}. 

Moreover, as shown in Fig.~\ref{fig:Fig3}(a), the longitudinal susceptibility exhibits algebraic divergence $\chi_h \propto T^{-7/4}$, while the transverse susceptibility shows logarithmic divergence $\chi_g \propto \ln T$, with the critical exponent $\alpha=0$. It is worth noting that the quantum supercritical scaling $\chi_h \propto T^{-\gamma/z\nu}$ typically dominates over the quantum critical one $\chi_g \propto T^{-\alpha/z\nu}$, since $\gamma > \alpha$ holds for conventional universality classes. This indicates that the longitudinal magnetic susceptibility provides a more sensitive probe for the QCP. 

\subsection{BKT QCP}

At $\Delta = -1$ and $g_c \approx 0.58$, the XXZ chain exhibits a BKT QCP, where the low-energy properties can be described by a sine-Gordon field theory~\cite{fradkin2013field}, 
\begin{equation}
    \mathcal{L}_{\rm BKT} = \frac{1}{2} (\partial_\mu \phi)^2 + u \cos(2\xi \phi) + h\cos(\xi \phi),
\end{equation}
where $\xi$ is a parameter, $u$ is the coupling strength and $h$ is the longitudinal field. In this case, the longitudinal field $h$ still couples to the FM order parameter and  breaks the U(1) symmetry, thus driving the quantum supercritical phenomena near the BKT QCP. According to the renormalization group analysis of the sine-Gordon field theory, the scaling dimension of the order parameter is $\Delta_O=1/2$, yielding that the quantum supercritical exponent is $\sigma =( d+z-\Delta_O)/z = 3/2$. Therefore, for the BKT QCP, the crossover lines of the QSR follow the $T\propto h^{2/3}$ scaling, as shown in Fig.~\ref{fig:Fig1}(c). 

\begin{figure*}[t!]
    \centering
    \includegraphics[width=1\linewidth]{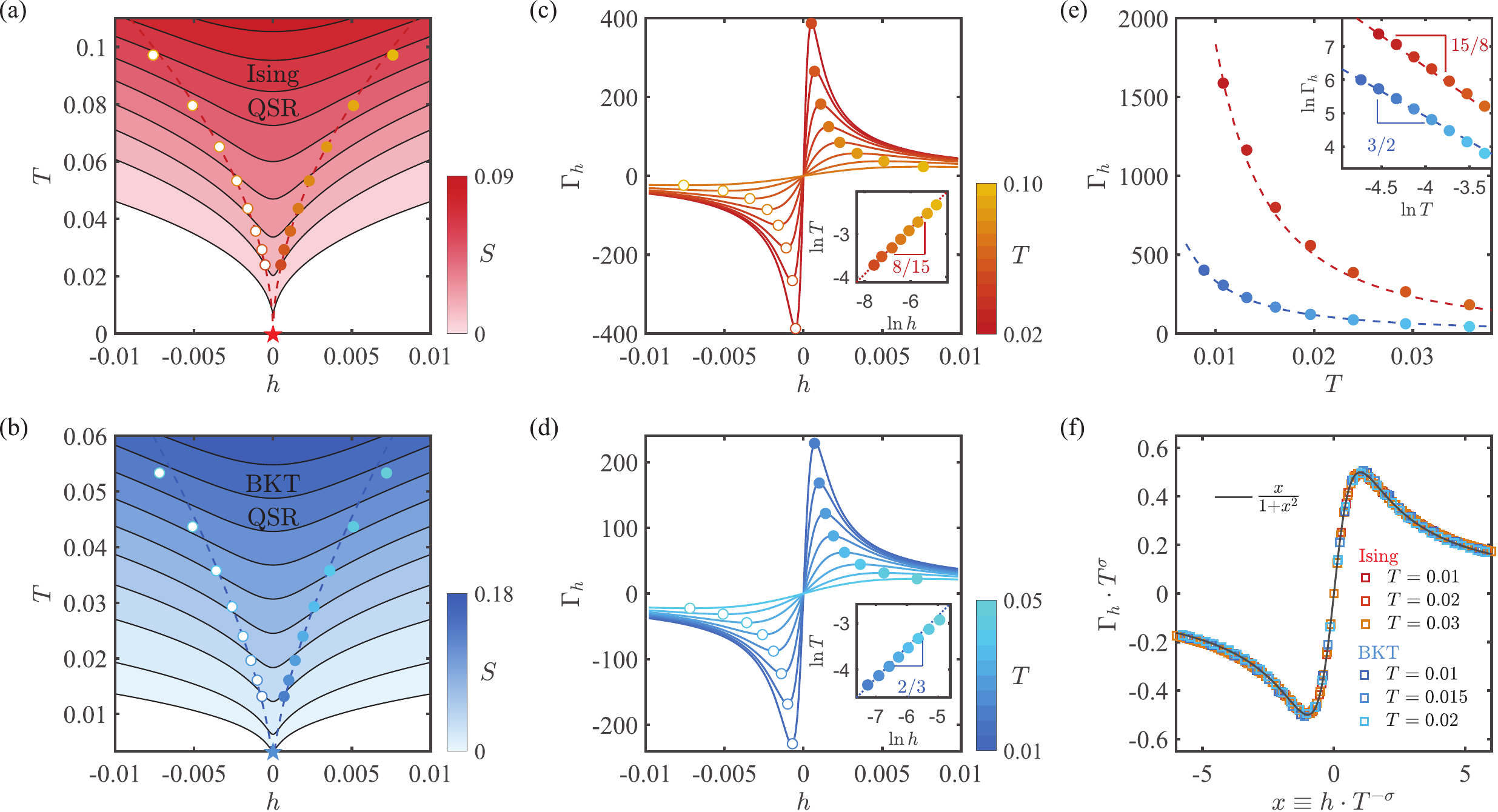}
    \caption{(a,b) Thermal entropy in the $h$-$T$ plane near the Ising QCP (red star, $\Delta = -1.5$, $g_c \approx 0.99$) and the BKT QCP (blue star, $\Delta = -1$, $g_c \approx 0.58$), respectively. Dashed lines represent the crossover lines of the QSR. (c,d) The Gr\"uneisen ratio $\Gamma_h$ driven by $h$ near these QCPs. Solid dots indicate the peak positions, while hollow dots indicate the dip positions. Insets illustrate the quantum supercritical law $T\propto h^{1/\sigma}$ of the crossover lines ($\sigma = 15/8$ for the Ising QCP and $\sigma = 3/2$ for the BKT QCP). (e) Peak values of the Gr\"uneisen ratio $\Gamma_h$. Inset shows the quantum supercritical scaling $\Gamma_h \propto T^{-\sigma}$ of these peak values. (f) Data collapse of the $\Gamma_h$ data for both the Ising case and the BKT case. {The scaling functions are rescaled by $1.4 \phi_\Gamma(1.8x) \rightarrow \phi_\Gamma(x)$ for the Ising case, and $1.5 \phi_\Gamma(2.1x)\rightarrow \phi_\Gamma (x)$ for the BKT case.} Black solid line represents the approximation $x/(1+x^2)$ of the scaling function of Gr\"uneisen ratio. }
    \label{fig:Fig4}
\end{figure*}

Then, the universal form given in Eq.~\ref{freeenergy} is no longer satisfied. The parameter $\tilde{g}$ opens the energy gap exponentially slow, leading to the logarithmic divergence of $\chi_g$ shown in Fig.~\ref{fig:Fig3}(b). The parameter $\tilde{g}$ has almost no impact on the quantum critical behavior, thus the two-variable hyperscaling function becomes a one-variable scaling function $\phi_f(x,y) \rightarrow \phi_f(x)$. In this case, the singular part of the free energy $F$ follows another universal form 
\begin{equation}
    F = T^{d/z+1} \phi_f\left( \frac{h}{T^{\sigma}} \right), 
    \label{freeenergy2}
\end{equation}
with $\sigma = 3/2$ for the (1+1)D BKT universality class. Near the BKT QCP, the longitudinal susceptibility follows a universal scaling $\chi_h \propto T^{-1}\ln T$, as shown in Fig.~\ref{fig:Fig3}(b). Then, the quantum-critical universal scaling of the specific heat near the QCP can be derived from Eq.~\ref{freeenergy}, which reads $C \equiv -T{\partial^2 F}/{\partial T^2} \propto T^{d/z}$. For both Ising and BKT QCP, the low-energy excitation is gapless with linear dispersion ($z=1$), leading to the linear low-temperature specific heat, as shown in Fig.~\ref{fig:Fig3}(c). 

\section{Quantum supercritical MCE}
\label{secIV}
In this section, we investigate the MCE induced by the longitudinal field $h$. Near a QCP, derived from the Eq.~\ref{freeenergy2}, the singular part of the thermal entropy also has a universal form 
\begin{equation}
    S = T^{d/z} \phi_s\left( \frac{h}{T^{\sigma}} \right), 
\end{equation}
where $\phi_s(x)$ is the scaling function of the entropy. It is worth noting that the entropy exhibits even parity with respect to the field $h$, implying that $\phi_s(x)$ is an even function. Figure~\ref{fig:Fig4}(a,b) illustrate the thermal entropy $S$ near the Ising QCP and the BKT QCP, respectively. As we can see, at a fixed temperature, the entropy reaches its maximum above QCPs. The longitudinal field $h$, which couples to the order parameter, rapidly drives the system into a spin-ordered phase with lower thermal entropy. Therefore, the isentropic lines in Fig.~\ref{fig:Fig4}(a,b) illustrate dips in the QSRs (specifically, above the QCPs), revealing the quantum-supercritical cooling effect during an adiabatic demagnetization process. Particularly, the $h$ field exhibits a stronger ability to regulate thermal entropy, i.e., the MCE, at the Ising QCP than at the BKT QCP. 

To measure the quantum supercritical MCE, we calculate the longitudinal Gr\"uneisen ratio $\Gamma_h\equiv 1/T(\partial T/\partial h)_S = - (\partial S/\partial h)_T/C$ near these two QCPs, as illustrated in Fig.~\ref{fig:Fig4}(c,d). At a fixed temperature, $\Gamma_h$ exhibits a peak-dip structure, with the peak and dip positions marked by solid and hollow dots, respectively. As the temperature decreases, the peaks and dips sharpen progressively. Notably, the peak and dip positions correspond to the crossover lines of the QSRs shown in Fig.~\ref{fig:Fig4}(a,b). These crossover lines obey the quantum supercritical scaling $T \propto h^{1/\sigma}$, with $\sigma = 15/8$ for the Ising QCP and $\sigma = 3/2$ for the BKT QCP [see insets of Fig.~\ref{fig:Fig4}(c,d)]. Furthermore, the peak and dip values of the Gr\"uneisen ratio diverge at low temperatures, following $\Gamma_h \propto T^{-\sigma}$, as shown in Fig~\ref{fig:Fig4}(e). Consistent with the entropy results, the Gr\"uneisen ratio near the Ising QCP exhibits a stronger divergence than that near the BKT QCP. 

Similar to other thermodynamics discussed above, near a QCP, the longitudinal Gr\"uneisen ratio $\Gamma_h$ also possesses a universal form 
\begin{equation}
    \Gamma_h = - \frac{(\partial S/\partial h)_T}{T(\partial S/\partial T)_h} = T^{-\sigma} \phi_{\Gamma} \left( \frac{h}{T^{\sigma}} \right), 
\end{equation}
where 
\begin{equation}
    \phi_{\Gamma} (x) = \frac{\phi_s'(x)}{\frac{d}{z} \phi_s(x) - \sigma x \phi_s'(x)}
\end{equation}
is the scaling function of the longitudinal Gr\"uneisen ratio. Therefore, we can collapse the $\Gamma_h$ data in different temperatures and fields onto a single curve by rescaling the axes. Remarkably, the $\Gamma_h$ data near both the Ising and BKT QCP is collapsed onto nearly the same scaling function $\phi_{\Gamma}(x)$, as shown in Fig.~\ref{fig:Fig4}(f). Data points in Fig.~\ref{fig:Fig4}(f) undergo coordinate scaling. For conventional universality classes, both the critical exponents and the scaling function are unique; however, the quantum supercritical scaling function $\phi_{\Gamma}(x)$ of the Gr\"uneisen ratio is more universal, which is even independent of the specific universality class. To understand this, we take the even parity of $h$ into consideration. If we expand the $\phi_s(x)$ to the second order $\phi_s(x) \approx A + B x^2$, then the scaling function of $\Gamma_h$ can be expressed as 
\begin{equation}
    \phi_{\Gamma}(x) \approx \frac{2Bzx}{Ad + B(d - 2z\sigma) x^2}. 
\end{equation}
By rescaling the axes as $\left(\frac{B}{A} - \frac{2B\sigma z}{Ad}\right)^{1/2}x \rightarrow x$ and $\frac{2z}{d} \left(\frac{A}{B} - \frac{2A\sigma z}{Bd}\right)^{1/2} \phi_{\Gamma}(x) \rightarrow \phi_{\Gamma}(x)$, consequently, we can obtain a very concise approximation $\phi_{\Gamma} (x) \approx x/(1+x^2)$, which can describe the quantum supercritical scaling function very well [see Fig.~\ref{fig:Fig4}(f)]. 

Then, this concise approximation implies some additional universal scaling laws of the quantum supercritical Gr\"uneisen ratio. For $h \ll T^{\sigma}$ (i.e., $x\ll 1$), the scaling function satisfies $\phi_{\Gamma} (x) \approx x$, indicating $\Gamma_h \propto T^{-2\sigma} h$. Therefore, above the QCP ($h=0$), the Gr\"uneisen ratio vanishes, reflecting the absence of the MCE. At a fixed field $h$, the Gr\"uneisen ratio diverges as $\Gamma_h \propto T^{-2\sigma}$ in the QSR, which is even faster than the $T^{-\sigma}$ scaling of the peak/dip values. For $h\gg T^{\sigma}$ (i.e., $x\gg 1$), the scaling function approaches $\phi_{\Gamma} (x) \approx 1/x$, implying $\Gamma_h \propto h^{-1}$ in the gapped phase. 

\section{Discussion}
\label{secV}
In this work, we utilize tensor-network methods to investigate the quantum supercriticality in the FM-AFM spin chain. In its ground-state phase diagram, we identify an Ising QCP and a BKT QCP, which belong to the (1+1)D Ising universality class and the (1+1)D BKT universality class, respectively. While previous studies have focused exclusively on the Ising quantum supercriticality~\cite{wang2025,Lv2025}, we reveal the quantum supercritical phenomena of a BKT QCP, extending the concept of quantum supercriticality. Near these two QCPs, we characterize a quantum supercritical scaling law of the longitudinal susceptibility $\chi_h \propto T^{-\gamma/z\nu}$, which is stronger than the quantum critical scaling of the transverse susceptibility $\chi_g \propto T^{-\alpha/z\nu}$ ($\gamma > \alpha$ for typical universality classes), providing a more sensitive probe of the QCP. 

Moreover, we reveal the quantum supercritical MCE of these two QCPs, with the Gr\"uneisen ratio following the scaling law $\Gamma_h \propto T^{-\sigma}$ ($\sigma = 15/8$ for the Ising QCP and $\sigma = 3/2$ for the BKT QCP). Strikingly, we propose a concise approximation for the quantum supercritical scaling function, $\phi_{\Gamma}(x) \approx x/(1+x^2)$, which is suitable for both the (1+1)D Ising and BKT universality classes, and maybe is even independent of universality classes. Generally, for a given universality class, not only the critical exponents, but also the scaling functions are unique. This approximation facilitates the study of the quantum supercritical MCE. 

For different spin models, whose QCPs belong to the same universality class, the scaling functions of the Gr\"uneisen ratio $\Gamma_g$ driven by quantum-fluctuation field $g$ coincide after appropriate rescaling of the axes~\cite{xuan2025}. Then, the quantum-critical scaling functions of $\Gamma_g$ are distinct for different universality classes and can therefore be used to distinguish QCPs. However, we find that for the quantum supercritical MCE, scaling functions of $\Gamma_h$ are nearly identical for the (1+1)D Ising and BKT universality classes. Consequently, the quantum supercritical MCE behavior can be obtained solely from the critical exponents, without needing to calculate a specific model. This universality originates from the symmetry associated with the symmetry-breaking field $h$ that drives the quantum supercritical phenomena, and extends beyond the universality of quantum phase transitions. Given this, it is important to calculate other quantum-supercritical scaling functions in the future, which will help establish a universal dictionary of such quantum supercritical phenomena. 

\begin{acknowledgements}
We are grateful to Yuan Gao and Junsen Wang for insightful discussions. This work was supported by the National Key Projects for Research and Development of China (Grant No.~2024YFA1409200), the National Natural Science Foundation of China (Grant Nos.~12534009, 12504186, and 12447101), and the Strategic Priority Research Program of Chinese Academy of Sciences (Grant No.~XDB1270100). We thank the HPC-ITP for the technical support and generous allocation of CPU time. 
\end{acknowledgements}

\bibliography{qsc}

\clearpage
\newpage
\onecolumngrid
\begin{center}
\textbf{\large{Appendix}}
\end{center}
\onecolumngrid

\subsection{Order parameters and the BKT critical field}

\begin{figure*}[h]
    \centering
    \includegraphics[width=1\linewidth]{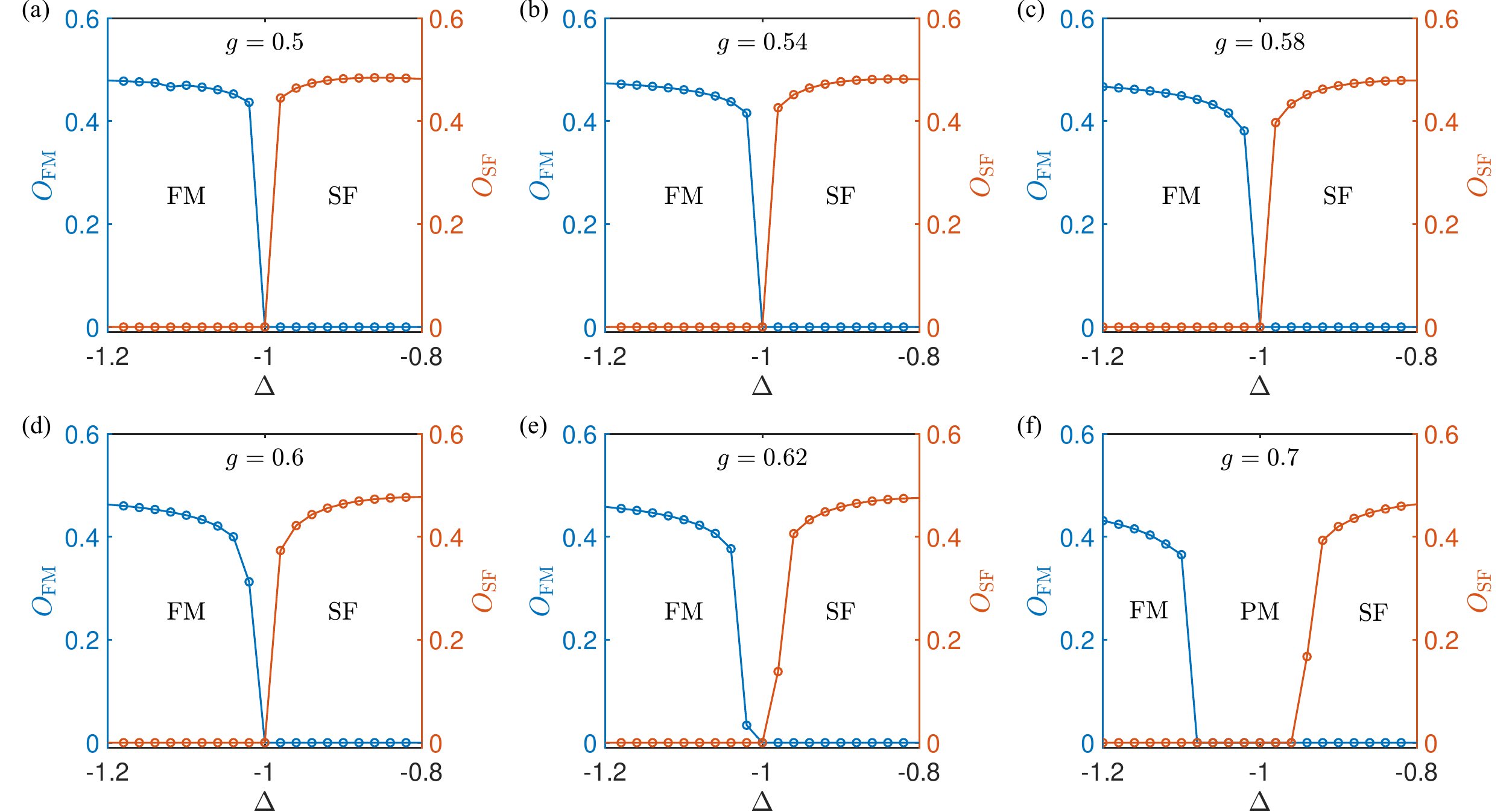}
    \caption{Order parameters of the $L=128$ spin chain calculated by DMRG method with the maximum bond dimension $D = 500$, under periodical boundary condition. $O_{\rm FM}$ is the ferromagnetic (FM) order parameter, while $O_{\rm SF}$ is the spin-flop (SF) order parameter. The finite-size critical field is about $g_c\approx 0.6$. (a,b) panels illustrate the $g < g_c$ case, and (c-f) panels exhibit the $g \geq g_c$ case.}
    \label{fig:Appendix_Fig1}
\end{figure*}

Here, we present the order parameters as functions of $\Delta$ for various transverse fields. $O_{\rm FM} \equiv \frac{1}{L}\sum_{i=1}^L\braket{S_i^z}$ denotes the order parameter of the ferromagnetic (FM) phase, while $O_{\rm SF} \equiv \frac{1}{L}\sum_{i=1}^L (-1)^i\braket{S_i^y}$ corresponds to the spin-flop (SF) phase. For the $L=128$ spin chain, the finite-size critical field is approximately $g_c \approx 0.58$. When $g<g_c$, as shown in panels (a,b) of Fig.~\ref{fig:Appendix_Fig1}, both order parameters exhibit clear jumps at $\Delta = -1$, indicating a first-order quantum phase transition. At $g = g_c$, the behavior of the order parameters suggests a second-order quantum phase transition [see panel (c) of Fig.~\ref{fig:Appendix_Fig1}]. For $g > g_c$, as illustrated in panels (d-f) of Fig.~\ref{fig:Appendix_Fig1}, the FM and SF phases separate, with the emergence of a paramagnetic (PM) phase. 

\clearpage
\newpage

\subsection{Connected correlation functions}

\begin{figure*}[h]
    \centering
    \includegraphics[width=1\linewidth]{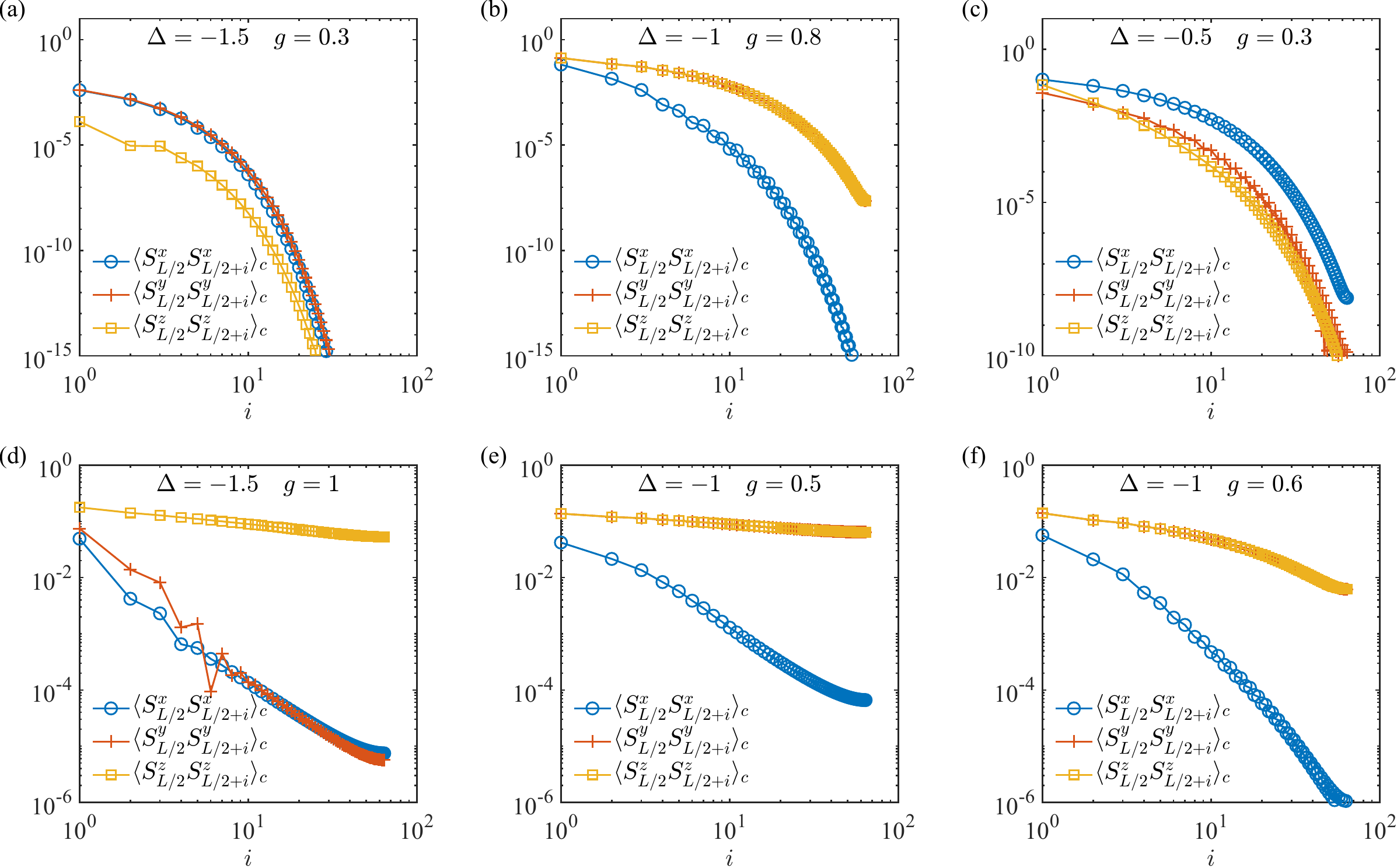}
    \caption{Connected correlation functions of the $L=128$ chain calculated by DMRG method with the maximum bond dimension $D = 500$, under periodical boundary condition. $\braket{ S^{\alpha}_{L/2}S^{\alpha}_{L/2+i}}_c \equiv \braket{ S^{\alpha}_{L/2}S^{\alpha}_{L/2+i}} - \braket{ S^{\alpha}_{L/2}}\braket{ S^{\alpha}_{L/2+i}}$ with $\alpha = x,y,z$ is the connected correlated function. 
    (a-c) panels illustrate three gapped cases, while (d-f) panels exhibit three gapless cases in the phase diagram of the XXZ chain.}
    \label{fig:Appendix_Fig2}
\end{figure*}

Here, we present the connected correlation functions $\braket{ S^{\alpha}_{L/2}S^{\alpha}_{L/2+i}}_c \equiv \braket{ S^{\alpha}_{L/2}S^{\alpha}_{L/2+i}} - \braket{ S^{\alpha}_{L/2}}\braket{ S^{\alpha}_{L/2+i}}$, with $\alpha = x,y,z$, in different phases. In the FM and SF phases, as shown in Fig.~\ref{fig:Appendix_Fig2}(a,c), the connected correlation functions decay exponentially, indicating the existence of a finite energy gap. For the $\Delta = -1$ case, the system is gapless with algebraic correlation in the $g<g_c$ region, and an energy gap opens gradually for $g>g_c$, as shown in Fig.~\ref{fig:Appendix_Fig2}(b,e). Additionally, the correlation length is infinite at the Ising QCP ($\Delta = -1.5$ and $g=1\approx g_c$) and the BKT QCP ($\Delta = -1$ and $g=0.58\approx g_c$), as shown in Fig.~\ref{fig:Appendix_Fig2}(d,f). 

\end{document}